\documentclass[aps,prd,onecolumn,superscriptaddress,showpacs,floatfix,10pt]{revtex4}
\usepackage{graphicx}
\usepackage{subfigure}
\usepackage{epsfig}

\begin{document}

\title{Plausible ``faster-than-light'' displacements in a two-sheeted spacetime}

\author{Fabrice Petit}
\email{f.petit@bcrc.be} \affiliation{Belgian Ceramic Research
Centre,\\4 avenue du gouverneur Cornez, B-7000 Mons, Belgium}

\author{Micha\"{e}l Sarrazin}
\email{michael.sarrazin@fundp.ac.be} \affiliation{Laboratoire de
Physique du Solide, Facult\'es Universitaires Notre-Dame de la Paix,
\\61 rue de Bruxelles, B-5000 Namur, Belgium}

\begin{abstract}
In this paper, we explore the implications of a two-point
discretization of an extra-dimension in a five-dimensional quantum
setup. We adopt a pragmatic attitude by considering the dynamics of
spin-half particles through the simplest possible extension of the
existing Dirac and Pauli equations. It is shown that the benefit of
this approach is to predict new physical phenomena while maintaining
the number of constitutive hypothesis at minimum. As the most
striking feature of the model, we demonstrate the possibility of
fermionic matter oscillations between the two four-dimensional
sections and hyper-fast displacements in case of asymmetric warping
(without conflicting special relativity). This result, similar to
previous reported ones in brane-world theories, is completely
original as it is derived by using quantum mechanics only without
recourse to general relativity and bulk geodesics calculation. The
model allows causal contact between normally disconnected regions.
If it proves to be physically founded, its practical aspects could
have deep implications for the search of extra-dimensions.
\end{abstract}

%


\pacs{11.10.Kk, 04.62.+v, 11.25.Wx}

\maketitle

\section{Introduction}

The idea that our observable universe could be a part of a more extended
N-dimensional spacetime (N$>$4) has a long tradition. It traces back to the
seminal work of Kaluza in 1921 who extended general relativity in five
dimensions in order to treat electromagnetism and gravitation on an equal
footing [1]. Unfortunately the model and its subsequent extensions reveal
unsuccessful in their aim of describing the physical reality such that the
multidimensional approach was abandoned for a while.

Since that time, much work has been carried out and modern
theoretical physics has led us to consider as more probable the
existence of a multidimensional universe. Recently, the idea was
reintroduced in the context of superstrings and brane-world
theories. This renewed interest can be explained by the fact that
multidimensional universes may adequately describe known forces and
particles and also explain the hierarchy between the gravitational
and electroweak scales [2,3]. These scenarios have been extended
much in recent years and there is now some approaches suggesting
that many connected parallel branes could exist in an extended bulk
(thus creating a so-called ``manyfold''). As summarized in Ref. [4],
the existence of such a multi-sheeted spacetime could shed some
light on a number of puzzling cosmological problems including the
nature of dark matter structures (which would be identical to normal
matter but located in a distinct sheet), as well as their
invisibility (the gauge fields and most notably electromagnetism
would be confined within the branes such that the structures
belonging to distinct branes would be mutually invisible).

Following this line of thought, other recent developments have tried to
improve the original Kaluza's idea of unification by considering discrete
extra-dimensions instead of continuous ones. In these models, each discrete
point in the fifth dimension is endowed with its own metric field such that
the resulting universe appears like a multi-sheeted spacetime composed by a
collection of parallel 4D sheets in interaction. Although such latticized
models are recognized to suffer from drawbacks (potentially curable), they
exhibit promising phenomenological properties, in particular in cosmology
[5,6].

Discrete extra-dimensions have also been studied in the spirit of
non commutative geometry. One of the most promising approaches is
that of Connes and Lott [7], Viet and Wali [8]. Their formulation
considers a product manifold comprising a continuous
four-dimensional part times a discrete one, typically $M_4\times
Z_2$. The non trivial feature of their theory is the coupling of the
two spacetime sheets which results in the Higgs field. In several
aspects, this construction appears as a reminiscence of the five
dimensional Kaluza-Klein model with the fifth dimension restricted
to only two points.

In recent papers, present authors have developed a phenomenological
model of a five dimensional two-sheeted spacetime and studied the
quantum behavior of massive particles in such a universe. Different
mathematical approaches were used involving either a non-commutative
product manifold [9] or a two-point discretization of the fifth
dimension [10]. In the latter approach, the compact extra dimension
was treated through a naive discretization scheme involving a finite
difference analysis. The procedure was inspired from
''multigravity'' theories where massive gravitons result as a
consequence of a latticized extra-dimension [5]. Extending such a
procedure to the case of the five dimensional Dirac equation was
demonstrated to provide several original results like fermionic
matter oscillations between the two four-dimensional sections
[9-12]. Therefore, despite all roughness of the approach, the
prediction of possible new quantum phenomena made the theory
interesting to study.

In the present paper, we propose to extend further the discussion of
Refs. [9-12] by considering an asymmetrically warped background
instead of a flat one. Since each four-dimensional section is now
endowed with its own warp factor, different physical length scales
can be defined on the sheets. We show that under some circumstances,
a massive particle can oscillate between the two four-dimensional
sections and exhibit hyperfast velocities from the perspective of a
four-dimensional observer thanks to the different length
scales in the two sheets. This result is similar to that obtained by Chung {%
and} Freese in Ref.[13] except that it is here derived by using quantum
mechanics exclusively without recourse to geodesic calculation.

The paper is structured as follows. In section II, we derive the
two-sheeted Dirac equation for the chosen asymmetrically warped
background. The solutions of the Dirac equation are given and
compared with the usual ones which are recovered at the decoupling
limit (infinite distance between the sheets). Then, in the third
part, we demonstrate the oscillatory behavior in the case of a
positive energy particle and the possibility of hyperfast motions.
In the fourth section, the non relativistic limit of the Dirac
equation is derived and the quantum dynamics is studied at low
energies. It is demonstrated that the oscillatory behavior and the
hyperfast velocities still survive in the non relativistic limit,
thus enabling a possible experimental confirmation of the model.

\section{Two-sheeted Dirac equation in a warped background}

\subsection{Mathematical framework}

In this part, we shall derive the two-sheeted Dirac equation. We start by
considering the usual 5D covariant Dirac equation:
\begin{equation}
\left( {i}\not{D}\text{ }{-}\text{ }{m}\right) \psi =i\gamma ^A(x,\lambda )%
\not{D}_A\psi -m\psi =0  \label{eq1}
\end{equation}
with $A\in \left\{ {0,1,2,3,4}\right\} $ and where $\gamma ^A(x,\lambda )$
are the curvature dependent Dirac matrices. $\not{D}_A$ is the covariant
derivative given by :
\begin{equation}
\not{D}_A=\partial _A+\Gamma _A(x,\lambda )  \label{eq2}
\end{equation}
with $\Gamma _A(x,\lambda )$ the spin connections and $\partial
_A=\partial /\partial x^A$. In Eq. (1) and (2), ``$x$'' refers to
the four-dimensional coordinates and $\lambda $ to the transverse
extra dimension (a subscript or an exponent equals to 4 corresponds
to $\lambda )$. Note that throughout this paper, we will work in
units where $c=1,$ $\not{h}=1$. Now, we assume that the five
dimensional metric takes the following specific form
\begin{equation}
ds^2=g_{AB}dx^Adx^B=dt^2-R^2(\lambda )\left[ {dx^2+dy^2+dz^2}\right]
+d\lambda ^2  \label{eq3}
\end{equation}
where $R\left( \lambda \right) $ is the warp factor. Note that we
consider that the warping only involves the space interval of the
line element and not the temporal part. This is similar to the
choice made by Chung and Freese in Ref. [13] and differs from the
classical brane-world approach where the metric takes a canonical
form. Our choice of metric corresponds to an asymmetric warping as
defined in Ref. [13]. Since the results of this paper are not
affected by the exact form of $R\left( \lambda \right) $, we are
keeping this general form throughout this paper. Note that a
signature $(+,-,-,-,+)$ is chosen instead of the usual $(+,-,-,-,-)$
one. It is straightforward to show that a timelike extra-dimension
is necessary to ensure energy conservation in the present model (a
spacelike extra-dimension leads to a non-hermitian Hamiltonian as it
is obvious from Eqs.(101) and (102) hereafter).

The five dimensional Dirac matrices in curved space take the form:
\begin{equation}
\gamma ^A(x,\lambda )=e_a^A(x,\lambda )\gamma ^a  \label{eq4}
\end{equation}
where $e_a^A$ define the vielbein according to:
\begin{equation}
g^{AB}=e_a^A(x)e_b^B(x)\eta^{ab}  \label{eq5}
\end{equation}
with $g_{AB}$ the 5D metric and $\eta_{ab}$ the five dimensional
metric tensor of the Minkowski spacetime. The vielbein is given by:
\begin{equation}
e_a^A(x)=diag\left( {1,\frac 1R,\frac 1R,\frac 1R,1}\right)  \label{eq6}
\end{equation}
which leads to :
\begin{equation}
\gamma ^0(x,\lambda )=\gamma ^0,\text{ }\gamma ^i(x,\lambda )=\frac 1R\gamma
^i,\text{ }\gamma ^4(x,\lambda )=\gamma ^5  \label{eq7}
\end{equation}
where $\gamma ^5=i\gamma ^0\gamma ^1\gamma ^2\gamma ^3$
anti-commutes with the usual Dirac matrices $\gamma ^\mu $ in flat
space, such that one verifies $\left\{ \gamma ^a,\gamma ^b\right\}
=2\eta ^{ab}$. The spin connection must satisfy the expression
\footnote{It is straightforward to show that the expression (8) of
the spin connection is equivalent to the more familiar one $ \Gamma
_A=\frac 18\omega _{abA}\left[ \gamma ^a,\gamma ^b\right] $ where
the connexion term is given by the usual expression $ \omega
_{abA}=\eta _{ac}\omega _{bA}^c=\eta _{ac}e_B^c\left( \frac
\partial {\partial x^A}e_b^B+\Gamma _{CA}^Be_b^C\right)
$. See for instance T. R. Slatyer, R. R. Volkas, ''Cosmology and
Fermion Confinement in a Scalar-Field-Generated Domain Wall Brane in
Five Dimensions'', JHEP 04 (2007) 062 (hep-ph/0609003) or S. L.
Parameswarana, S. Randjbar-Daemib, A. Salvioa, ''Gauge Fields,
Fermions and Mass Gaps in 6D Brane Worlds'', Nucl.Phys. B767 (2007)
54-81 (hep-th/0608074)}:
\begin{equation}
\Gamma _A(x,\lambda )=\frac 14\gamma _B\left[ {\partial _A\gamma
^B(x,\lambda )+\Gamma _{CA}^B\gamma ^C(x,\lambda )}\right]  \label{eq8}
\end{equation}
where $\Gamma _{CA}^B$ are the Christoffel symbols for the metric field
defined above. From (8), the following covariant derivative terms can be
found :
\begin{equation}
\not{D}_0=\partial _0,\not{D}_4=\partial _4,\not{D}_i=\partial _i-\frac 12%
\dot{R}\gamma ^i\gamma ^5  \label{eq9}
\end{equation}
where a dot implies a derivation along the fifth dimension and the gamma
matrices in Eq.(9) are those of flat space. The five dimensional Dirac
equation for the metric field (3) finally becomes :
\begin{equation}
\left\{ {i\gamma ^0\partial _0+\frac iR\gamma ^i\partial _i+i\gamma
^5\partial _4+i\frac 32\frac{\dot{R}}R\gamma ^5-m}\right\} \psi =0
\label{eq10}
\end{equation}
Let us now discretize the fifth dimension. The way to proceed is
inspired by those described in Refs. [5-6]. However, instead of
considering a one dimensional lattice, we focus on a more restricted
compact extra-dimension containing two points located at coordinates
$\lambda \in \left\{ {-\delta /2,+\delta /2}\right\} $ (see Ref.
[10] where such an approach had been
developed). At each site, there is thus a four-dimensional submanifold $%
X_{\pm }$ where the particle wave function takes the local form
$\psi (x,\lambda =\pm \delta /2)=\psi _{\pm }(x)$. In the proposed
geometrical framework the derivative $\partial _4$ along the
discrete dimension simply reduces to a finite difference [10] (see
Appendix A) :

\begin{equation}
\left. {\partial _4\psi }\right| _{\pm }\to g\left( {\psi _{\pm }-\psi _{\mp
}}\right) \text{ and }\dot{R}_{\pm }=g\left( {R_{\pm }-R_{\mp }}\right)
\label{eq11}
\end{equation}
where $g$ is the inverse of the distance $\delta $ between the
four-dimensional sections and $R_{-}=R(\lambda =-\delta /2)$, $%
R_{+}=R(\lambda =+\delta /2)$ are the ''projected'' warp factors on
the two sheets.

Using those expressions, the five dimensional Dirac equation breaks down
into a set of two coupled four-dimensional like Dirac equations:
\begin{equation}
i\gamma ^0\partial _0\psi _{+}+\frac i{R_{+}}\gamma ^i\partial _i\psi
_{+}+i\gamma ^5g\left( {\psi _{+}-\psi _{-}}\right) +i\gamma ^5\frac
3{2R_{+}}g\left( {R_{+}-R_{-}}\right) \psi _{+}-m\psi _{+}=0  \label{eq12}
\end{equation}
\begin{equation}
i\gamma ^0\partial _0\psi _{-}+\frac i{R_{-}}\gamma ^i\partial _i\psi
_{-}+i\gamma ^5g\left( {\psi _{-}-\psi _{+}}\right) +i\gamma ^5\frac
3{2R_{-}}g\left( {R_{-}-R_{+}}\right) \psi _{-}-m\psi _{-}=0  \label{eq13}
\end{equation}
This equation can be rewritten in a compact matrix form such that
\begin{equation}
\left( {i\not{D}-m}\right) \psi =0  \label{eq14}
\end{equation}
with
\begin{equation}
{\not{D}=\Gamma ^0\ \partial _0+\frac 1R\Gamma ^\eta \ \partial _\eta +g\
\Gamma ^5+g\ \Delta }  \label{eq15}
\end{equation}
provided that
\begin{equation}
\psi =\left( {{\
\begin{array}{c}
{\psi _{+}} \\
{\psi _{-}}
\end{array}
}}\right) ,\Gamma ^0=\left( {{\
\begin{array}{cc}
{\gamma ^0} & 0 \\
0 & {\gamma ^0}
\end{array}
}}\right) ,\frac 1R=\left( {{\
\begin{array}{cc}
{1/R_{+}} & 0 \\
0 & {1/R_{-}}
\end{array}
}}\right) ,\Gamma ^\eta =\left( {{\
\begin{array}{cc}
{\gamma ^\eta } & 0 \\
0 & {\gamma ^\eta }
\end{array}
}}\right) ,  \label{eq16}
\end{equation}
\begin{equation}
\Gamma ^5=\left( {{\
\begin{array}{cc}
{\gamma ^5} & {-\gamma ^5} \\
{-\gamma ^5} & {\gamma ^5}
\end{array}
}}\right) ,\Delta =\frac 32\ \left( {{\
\begin{array}{cc}
{\gamma ^5\ \left( {R_{+}-R_{-}}\right) /R_{+}} & 0 \\
0 & {\gamma ^5\ \left( {R_{-}-R_{+}}\right) /R_{-}}
\end{array}
}}\right) .  \label{eq17}
\end{equation}
Eq. (14) is the two-sheeted Dirac equation for the metric field (3) which
will be studied in this paper. Note that it takes a form which reminds the
usual Dirac equation except for the last two terms $ig\Gamma _5$ and $%
ig\Delta $ which generate the coupling between the sheets. It can be noted
that no interaction occurs anymore at the limit where the sheets are
infinitely separated, i.e. $g\to 0$.

\subsection{Asymmetrical two-sheeted Klein-Gordon equation}

To solve Eq.(14) it is mandatory to introduce the auxiliary field $\phi $
given by :
\begin{eqnarray}
\psi &=&\left( {i\not{D}+m}\right) \phi  \label{eq18} \\
&=&\left( {i\Gamma ^0\ \partial _0+i\ \frac 1R\Gamma ^i\ \partial _i+m+i\ g\
\Gamma ^5+i\ g\ \Delta }\right) \phi  \nonumber
\end{eqnarray}
Then, it can be shown that $\phi $ is solution of:
\begin{equation}
\left( {i\not{D}-m}\right) \left( {i\not{D}+m}\right) \phi =\left\{ {%
-\partial _0^2+\frac 1{R^2}\partial _i^2-m^2-g\ \mathrm{H}^i\ \partial
_i-g^2\ \Gamma }\right\} \ \phi =0  \label{eq19}
\end{equation}
with:
\begin{equation}
\mathrm{H}^i=\left( {{\
\begin{array}{cc}
0 & {-\gamma ^i\gamma ^5\left( {1/R_{+}-1/R_{-}}\right) } \\
{-\gamma ^i\gamma ^5\left( {1/R_{-}-1/R_{+}}\right) } & 0
\end{array}
}}\right) ,\text{ }\Gamma =\left( {{\
\begin{array}{cc}
{\Sigma _{+}} & {-\Pi } \\
{-\Pi } & {\Sigma _{-}}
\end{array}
}}\right)  \label{eq20}
\end{equation}
where we have set :
\begin{equation}
\Sigma _{+}=\frac{29R_{+}^2-30R_{+}R_{-}+9R_{-}^2}{4R_{+}^2},\text{ }\Sigma
_{-}=\frac{29R_{-}^2-30R_{+}R_{-}+9R_{+}^2}{4R_{-}^2},\text{ }\Pi =2-\frac 32%
\frac{\left( {R_{+}-R_{-}}\right) ^2}{R_{+}R_{-}}  \label{eq21}
\end{equation}
Following the usual procedure, we can solve this two-sheeted Klein-Gordon
equation by using an ansatz of the form $\phi =\phi _0e^{-i\left( {%
E_pt-p\cdot x}\right) }$. Introducing this ansatz into Eq.(19) leads to the
system:
\begin{equation}
\left( {{\
\begin{array}{cc}
{E_p^2-\frac{p^2}{R_{+}^2}-m^2-g^2\Sigma _{+}} & {g^2\Pi -ig\left( {%
1/R_{+}-1/R_{-}}\right) \gamma ^5\gamma ^ip_i} \\
{g^2\Pi -ig\left( {1/R_{-}-1/R_{+}}\right) \gamma ^5\gamma ^ip_i} & {E_p^2-%
\frac{p^2}{R_{-}^2}-m^2-g^2\Sigma _{-}}
\end{array}
}}\right) \phi _0=0  \label{eq22}
\end{equation}

\subsection{Free-field solutions of the two-sheeted Klein-Gordon and Dirac
equations}

Let us determine the solutions of the two-sheeted Klein-Gordon and Dirac
equations. To simplify notations in the forthcoming results, it is
convenient to write:

\begin{equation}
\chi =\frac{R_{+}^2+R_{-}^2}{R_{+}^2R_{-}^2}\text{ and }\Xi =\alpha
^2p^4+\beta p^2+\gamma  \label{eq23}
\end{equation}
with
\begin{equation}
\alpha =\frac{R_{+}^2-R_{-}^2}{R_{+}^2R_{-}^2},\text{ }\beta =2g^2\left( {%
\frac 1{R_{-}}-\frac 1{R_{+}}}\right) \left[ {2\left( {\frac 1{R_{-}}-\frac
1{R_{+}}}\right) -\left( {\frac 1{R_{-}}+\frac 1{R_{+}}}\right) \left( {%
\Sigma _{+}-\Sigma _{-}}\right) }\right] ,  \label{eq24}
\end{equation}
\begin{equation}
\gamma =g^4\left[ {4\Pi ^2+\left( {\Sigma _{+}-\Sigma _{-}}\right) ^2}%
\right] ,  \label{eq25}
\end{equation}
\begin{equation}
\kappa =g^2\Pi ,\text{ }\tau =ig\left( {\frac 1{R_{-}}-\frac 1{R_{+}}}%
\right) .  \label{eq26}
\end{equation}

From Eq.(22), the energy eigenvalues $E$ and eigenvectors $\phi _0$ are
easily found:
\begin{equation}
E=E_p=\left\{ {m^2+p^2\frac \chi 2+g^2\frac{\left( {\Sigma _{+}+\Sigma _{-}}%
\right) }2+\frac{\sqrt{\Xi }}2}\right\} ^{1/2}\text{ with }\phi _0=\left( {{%
\
\begin{array}{c}
{\eta \phi _\lambda } \\
{\vartheta _\lambda \phi _\lambda }
\end{array}
}}\right)  \label{eq27}
\end{equation}
and
\begin{equation}
E=-E_p=-\left\{ {m^2+p^2\frac \chi 2+g^2\frac{\left( {\Sigma _{+}+\Sigma _{-}%
}\right) }2+\frac{\sqrt{\Xi }}2}\right\} ^{1/2}\text{ with }\phi _0=\left( {{%
\
\begin{array}{c}
{\eta \varphi _\lambda } \\
{\vartheta _\lambda ^{*}\varphi _\lambda }
\end{array}
}}\right)  \label{eq28}
\end{equation}
with $\eta =\frac 12\left[ {\alpha p^2-g^2\left( {\Sigma _{+}-\Sigma _{-}}%
\right) -\sqrt{\Xi }}\right] $ and

\begin{equation}
E=\tilde{E}_p=\left\{ {m^2+p^2\frac \chi 2+g^2\frac{\left( {\Sigma
_{+}+\Sigma _{-}}\right) }2-\frac{\sqrt{\Xi }}2}\right\} ^{1/2}\text{ with }%
\phi _0=\left( {{\
\begin{array}{c}
{\tilde{\eta}\phi _\lambda } \\
{\vartheta _\lambda \phi _\lambda }
\end{array}
}}\right)  \label{eq29}
\end{equation}
and
\begin{equation}
E=-\tilde{E}_p=-\left\{ {m^2+p^2\frac \chi 2+g^2\frac{\left( {\Sigma
_{+}+\Sigma _{-}}\right) }2-\frac{\sqrt{\Xi }}2}\right\} ^{1/2}\text{ with }%
\phi _0=\left( {{\
\begin{array}{c}
{\tilde{\eta}\varphi _\lambda } \\
{\vartheta _\lambda ^{*}\varphi _\lambda }
\end{array}
}}\right)  \label{eq30}
\end{equation}
with $\tilde{\eta}=\frac 12\left[ {\alpha p^2-g^2\left( {\Sigma
_{+}-\Sigma _{-}}\right) +\sqrt{\Xi }}\right] $. $\lambda =\pm \frac
12$ and refers throughout the paper to the helicity states of the
particle. We also have:
\begin{equation}
\vartheta _\lambda =\kappa +2\tau \lambda p\text{ with }\phi _\lambda
=\left( {{\
\begin{array}{c}
{\chi _\lambda } \\
0
\end{array}
}}\right)  \label{eq31}
\end{equation}
and
\begin{equation}
\vartheta _\lambda ^{*}=\kappa -2\tau \lambda p\text{ with }\varphi _\lambda
=\left( {{\
\begin{array}{c}
0 \\
{\chi _\lambda }
\end{array}
}}\right)  \label{eq32}
\end{equation}
where $\chi _\lambda $ is the spinor such that, if one considers the
impulsion $\mathbf{p}$, it verifies the usual equation:
\begin{equation}
\sigma ^ip_i\chi _\lambda =2\lambda p\chi _\lambda  \label{eq33}
\end{equation}
One notes that $\left\{ \vartheta _\lambda ,\phi _\lambda \right\} $ and $%
\left\{ \vartheta _\lambda ^{*},\varphi _\lambda \right\} $ are the
eigensolutions of the operator $T$ given by:
\begin{equation}
T={\kappa -\tau \gamma ^5\gamma ^ip_i}=\left( {{\
\begin{array}{cc}
{\kappa +\tau \sigma ^ip_i} & 0 \\
0 & {\kappa -\tau \sigma ^ip_i}
\end{array}
}}\right)  \label{eq34}
\end{equation}
and which constitutes the non-diagonal terms of the matrix in Eq.
(22).

Using Eq.(18), it can be shown that:
\begin{equation}
\psi =\left( {{\
\begin{array}{cccc}
{E+m} & {-\frac{\sigma ^ip_i}{R_{+}}+ig\Theta _{+}} & 0 & {-ig} \\
{\frac{\sigma ^ip_i}{R_{+}}+ig\Theta _{+}} & {-E+m} & {-ig} & 0 \\
0 & {-ig} & {E+m} & {-\frac{\sigma ^ip_i}{R_{-}}+ig\Theta _{-}} \\
{-ig} & 0 & {\frac{\sigma ^ip_i}{R_{-}}+ig\Theta _{-}} & {-E+m}
\end{array}
}}\right) \phi _0  \label{eq35}
\end{equation}
with
\begin{equation}
\Theta _{\pm }={1+\frac 32\frac{R_{\pm }-R_{\mp }}{R_{\pm }}}  \label{eq36}
\end{equation}
For the positive energies, Eqs.(27) and (29) suggest to look for solutions
of the form:

\begin{equation}
\phi _0=\eta \left[ {{\
\begin{array}{c}
{\chi _\lambda } \\
0 \\
0 \\
0
\end{array}
}}\right] +\vartheta _\lambda \left[ {{\
\begin{array}{c}
0 \\
0 \\
{\chi _\lambda } \\
0
\end{array}
}}\right] \text{ or }\phi _0=\tilde{\eta}\left[ {{\
\begin{array}{c}
{\chi _\lambda } \\
0 \\
0 \\
0
\end{array}
}}\right] +\vartheta _\lambda \left[ {{\
\begin{array}{c}
0 \\
0 \\
{\chi _\lambda } \\
0
\end{array}
}}\right]  \label{eq37}
\end{equation}
Similarly for negative energies, Eqs.(28) and (30) suggest to look for
solutions of the form:

\begin{equation}
\phi _0=\eta \left[ {{\
\begin{array}{c}
0 \\
{\chi _\lambda } \\
0 \\
0
\end{array}
}}\right] +\vartheta _\lambda ^{*}\left[ {{\
\begin{array}{c}
0 \\
0 \\
0 \\
{\chi _\lambda }
\end{array}
}}\right] \text{ or }\phi _0=\tilde{\eta}\left[ {{\
\begin{array}{c}
0 \\
{\chi _\lambda } \\
0 \\
0
\end{array}
}}\right] +\vartheta ^{*}\left[ {{\
\begin{array}{c}
0 \\
0 \\
0 \\
{\chi _\lambda }
\end{array}
}}\right]  \label{eq38}
\end{equation}
Inserting the ansatz (37) and (38) into Eq.(35) leads (after normalization)
to the following solutions for the Dirac equation:

For $E=\pm E_p$
\begin{equation}
u_\lambda \left( p\right) =\frac 1{\sqrt{C}}\left\{ {{\
\begin{array}{c}
{\eta \left( {E_p+m}\right) \chi _\lambda } \\
{\left( {\eta \left( {\frac{2\lambda p}{R_{+}}+ig\Theta _{+}}\right)
-ig\vartheta _\lambda }\right) \chi _\lambda } \\
{\vartheta _\lambda \left( {E_p+m}\right) \chi _\lambda } \\
{\left( {\vartheta _\lambda \left( {\frac{2\lambda p}{R_{-}}+ig\Theta _{-}}%
\right) -ig\eta }\right) \chi _\lambda }
\end{array}
}}\right\} ,\text{ }v_\lambda \left( p\right) =\frac 1{\sqrt{C}}\left\{ {{\
\begin{array}{c}
{\left( {\eta \left( {\frac{2\lambda p}{R_{+}}-ig\Theta _{+}}\right)
+ig\vartheta _\lambda ^{*}}\right) \chi _\lambda } \\
\eta {\left( {E_p+m}\right) \chi _\lambda } \\
{\left( {\vartheta _\lambda ^{*}\left( {\frac{2\lambda p}{R_{-}}-ig\Theta
_{-}}\right) +ig\eta }\right) \chi _\lambda } \\
{\vartheta _\lambda ^{*}\left( {E_p+m}\right) \chi _\lambda }
\end{array}
}}\right\}  \label{eq39}
\end{equation}
where $u_\lambda (p)$ and $v_\lambda (p)$ refer respectively to the positive
and negative energy solutions.

For $E=\pm \tilde{E}_p$
\begin{equation}
\tilde{u}_\lambda \left( p\right) =\frac 1{\sqrt{\tilde{C}}}\left\{ {{\
\begin{array}{c}
{\tilde{\eta}\left( {\tilde{E}_p+m}\right) \chi _\lambda } \\
{\left( {\tilde{\eta}\left( {\frac{2\lambda p}{R_{+}}+ig\Theta _{+}}\right)
-ig\vartheta _\lambda }\right) \chi _\lambda } \\
{\vartheta _\lambda \left( {\tilde{E}_p+m}\right) \chi _\lambda } \\
{\left( {\vartheta _\lambda \left( {\frac{2\lambda p}{R_{-}}+ig\Theta _{-}}%
\right) -ig\tilde{\eta}}\right) \chi _\lambda }
\end{array}
}}\right\} ,\text{ }\tilde{v}_\lambda \left( p\right) =\frac 1{\sqrt{\tilde{C%
}}}\left\{ {{\
\begin{array}{c}
{\left( {\tilde{\eta}\left( {\frac{2\lambda p}{R_{+}}-ig\Theta _{+}}\right)
+ig\vartheta _\lambda ^{*}}\right) \chi _\lambda } \\
{\tilde{\eta}\left( {\tilde{E}_p+m}\right) \chi _\lambda } \\
{\left( {\vartheta _\lambda ^{*}\left( {\frac{2\lambda p}{R_{-}}-ig\Theta
_{-}}\right) +ig\tilde{\eta}}\right) \chi _\lambda } \\
{\vartheta _\lambda ^{*}\left( {\tilde{E}_p+m}\right) \chi _\lambda }
\end{array}
}}\right\}  \label{eq40}
\end{equation}

where, as previously, $\tilde{u}_\lambda (p)$ and $\tilde{v}_\lambda (p)$
refer respectively to the positive and negative energy solutions. In this
model the status of the negative energy is the same that in the classical
Dirac equation. The normalization constants $C$ and $\tilde{C}$ are easily
calculated:
\begin{equation}
C=\eta ^2\left\{ {\left( {E_p+m}\right) ^2+\frac{p^2}{R_{+}^2}+g^2\left( {%
1+\Theta _{+}^2}\right) }\right\} +\left| {\theta _\lambda }\right|
^2\left\{ {\left( {E_p+m}\right) ^2+\frac{p^2}{R_{-}^2}+g^2\left( {1+\Theta
_{-}^2}\right) }\right\} -2\eta \left| {\theta _\lambda }\right| ^2
\label{eq41}
\end{equation}
and
\begin{equation}
\tilde{C}=\tilde{\eta}^2\left\{ {\left( {\tilde{E}_p+m}\right) ^2+\frac{p^2}{%
R_{+}^2}+g^2\left( {1+\Theta _{+}^2}\right) }\right\} +\left| {\theta
_\lambda }\right| ^2\left\{ {\left( {\tilde{E}_p+m}\right) ^2+\frac{p^2}{%
R_{-}^2}+g^2\left( {1+\Theta _{-}^2}\right) }\right\} -2\tilde{\eta}\left| {%
\theta _\lambda }\right| ^2  \label{eq42}
\end{equation}
where $\left| {\vartheta _\lambda }\right| ^2=\kappa ^2+\xi ^2p^2$ with $\xi
=g\left( {\frac 1{R_{-}}-\frac 1{R_{+}}}\right) $.

It can be easily checked that at the limit where $R_{\pm }\to 1$, i.e. when
both sheets have identical warp factors, the above solutions conform to that
of the paper [10].

Reciprocally, at the decoupling limit (i.e. when $g\to 0$), the solutions
(39) and (40) become ``two times'' the classical Dirac ones expressed using
the coordinates ''$x$''. A careful calculation shows that the spinors of
positive energy become:

\begin{equation}
u_\lambda =\frac 1{\sqrt{2E_p\left( {E_p+m}\right) }}\left[ {{\
\begin{array}{c}
{\left( {E_p+m}\right) \chi _\lambda } \\
{\frac{2\lambda p}{R_{+}}\chi _\lambda } \\
0 \\
0
\end{array}
}}\right] \text{ and }\tilde{u}_\lambda =\frac 1{\sqrt{2\tilde{E}_p\left( {%
\tilde{E}_p+m}\right) }}\left[ {{\
\begin{array}{c}
0 \\
0 \\
{\left( {\tilde{E}_p+m}\right) \chi _\lambda } \\
{\frac{2\lambda p}{R_{-}}}
\end{array}
}}\right]  \label{eq43}
\end{equation}
while the energies become :

\begin{equation}
E_p=\sqrt{m^2+p^2\frac 1{R_{+}^2}}\text{ and }\tilde{E}_p=\sqrt{m^2+p^2\frac
1{R_{-}^2}}  \label{eq44}
\end{equation}
These solutions describe two four-dimensional spinors living in
distinct
sheets provided that the physical momenta $p_{+}$ and $p_{-}$ are $%
p_{+}=p/R_{+}$ and $p_{-}=p/R_{-}$ on the $(+)$ and $(-)$ sheet respectively.

This choice implies that the physical length coordinates are $x_{+}=R_{+}x$
and $x_{-}=R_{-}x$ on the $(+)$ and $(-)$ sheet respectively. Note that
under such a coordinate rescaling, both action and phase are conserved
since:
\begin{equation}
\mathbf{p\cdot x}=\mathbf{p}_{+}\cdot \left( {R_{+}}\mathbf{x}\right) =%
\mathbf{p}_{-}\cdot \left( {R_{-}}\mathbf{x}\right) =\mathbf{p}_{+}\cdot
\mathbf{x}_{+}=\mathbf{p}_{-}\mathbf{\cdot x}_{-}  \label{eq45}
\end{equation}

More clarifications on the coordinate rescaling will be given later in
section IV. Throughout this paper, $x$ and $p$ will refer to ``global''
coordinates and momenta whilst $x_{\pm }$ and $p_{\pm }$ will refer to
``physical'' or ``ordinary'' coordinates and momenta.

\section{Fermionic oscillations between the sheets and hyperfast
displacements}

\subsection{Particle motion between the two four-dimensional sections}

The predictions of the model can be illustrated by studying the following
state corresponding to an unpolarized particle of positive energy:
\begin{equation}
\psi =\frac 1{2\sqrt{V}}\left( {u_{1/2}e^{ip\cdot x}e^{-iE_pt}-\tilde{u}%
_{1/2}e^{ip\cdot x}e^{-i\tilde{E}_pt}}\right) +\frac 1{2\sqrt{V}}\left( {%
u_{-1/2}e^{ip\cdot x}e^{-iE_pt}-\tilde{u}_{-1/2}e^{ip\cdot x}e^{-i\tilde{E}%
_pt}}\right)  \label{eq46}
\end{equation}
By virtue of the two-sheeted structure of spacetime, $\psi $ is a
8-component spinor whose first fourth components are located on the $(+)$
sheet and the last fourth on the $(-)$ sheet.

The probability $P_{+}$ (resp. $P_{-})$ to find the particle in the $(+)$
sheet (resp. $(-)$ sheet) is simply given by the integration of $\left| {%
\psi _{+}}\right| ^2$(resp. $\left| {\psi _{-}}\right| ^2)$, the square norm
of the first (resp. last) fourth components of $\psi $, over the space
coordinates on the volume $V$. We get
\begin{equation}
P_{+}=A-2B\cos \left( {E_p-\tilde{E}_p}\right) t  \label{eq47}
\end{equation}
and
\begin{equation}
P_{-}=1-P_{+}  \label{eq48}
\end{equation}
with
\begin{equation}
\begin{array}{l}
A=\frac 1{2C}\left\{ {\eta ^2\left[ {\left( {E_p+m}\right) ^2+\frac{p^2}{%
R_{+}^2}+g^2\Theta _{+}^2}\right] +g^2\left| \vartheta \right| ^2+2g\eta
\left[ {\xi \frac{p^2}{R_{+}}-g\kappa \Theta _{+}}\right] }\right\} \\
+\frac 1{2\tilde{C}}\left\{ {\tilde{\eta}^2\left[ {\left( {\tilde{E}_p+m}%
\right) ^2+\frac{p^2}{R_{+}^2}+g^2\Theta _{+}^2}\right] +g^2\left| \vartheta
\right| ^2+2g\tilde{\eta}\left[ {\xi \frac{p^2}{R_{+}}-g\kappa \Theta _{+}}%
\right] }\right\}
\end{array}
\label{eq49}
\end{equation}
and
\begin{equation}
B=\frac 1{2\sqrt{C\tilde{C}}}\left\{ {\eta \tilde{\eta}\left[ {\left( {E_p+m}%
\right) \left( {\tilde{E}_p+m}\right) +\frac{p^2}{R_{+}^2}+g^2\Theta _{+}^2}%
\right] +g^2\left| \vartheta \right| ^2+2g\left[ {\eta +\tilde{\eta}}\right]
\left[ {\xi \frac{p^2}{R_{+}}-g\kappa \Theta _{+}}\right] }\right\}
\label{eq50}
\end{equation}
The presence of the cosine term in Eq.(47) shows that the particle
oscillates between the two sheets with a frequency $\nu $ proportional to
the difference of the two energy eigenvalues, i.e.:
\begin{equation}
2\pi \nu =\Delta E=\left| {E_p-\tilde{E}_p}\right|  \label{eq51}
\end{equation}
If one assumes that the mass term prevails over the other contributions,
then the frequency is given by (at first order):
\begin{equation}
2\pi \nu \cong \frac{\sqrt{\Xi }}{2m}  \label{eq52}
\end{equation}
and it can be easily checked that the oscillation frequency increases with
the coupling strength between the two sheets and it is enhanced for low mass
particles. At the limit of very low impulsion and for identical warp factors
($R_{+}=R_{-}$), we get $\nu =g^2/\left( {\pi m}\right) $ which is the
result found in Ref. [10] in a flat background.

Eq.(47) tells us that when the particle oscillates, it can disappear
periodically from the perspective of any four-dimensional observer.
This result is just the discrete counterpart of motions through the
bulk predicted in some brane-world theories. A common feature of all
those models (be they involving continuous or discrete
extra-dimension [9-12,14,15]) is the apparent violation of the
energy conservation from a four-dimensional point of view. It is
obvious that such a violation is only an artifact of low
dimensionality.

\subsection{Asymmetrical warp factors and velocities}

In order to illustrate the incidence of the different physical length scales
on the two sheets ($x_{+}\neq x_{-}$ in the more general case for a same
value of $x$), it is convenient to rewrite the equation (14) by multiplying
it by $\Gamma ^0$. Then, the two-sheeted Dirac equation can be recast as
\begin{equation}
i\frac \partial {\partial t}\psi =H\psi  \label{eq53}
\end{equation}
where $H$ is the two-sheeted Hamiltonian. In this form, the global velocity
operator $V$ can be trivially calculated:
\begin{equation}
\left( V\right) _i=\left( {\frac{\partial H}{\partial \mathbf{p}}}\right)
_i=\left( {{\
\begin{array}{cc}
{\frac 1{R_{+}}\gamma ^0\gamma ^i} & 0 \\
0 & {\frac 1{R_{-}}\gamma ^0\gamma ^i}
\end{array}
}}\right)  \label{eq54}
\end{equation}
with $\mathbf{p}$ the global momentum operator. The global velocity can then
be expressed:
\begin{equation}
v=\left\langle \psi \right| V\left| \psi \right\rangle =\frac
1{R_{+}}\left\langle {\psi _{+}}\right| \gamma ^0\gamma ^ie_i\left| {\psi
_{+}}\right\rangle +\frac 1{R_{-}}\left\langle {\psi _{-}}\right| \gamma
^0\gamma ^ie_i\left| {\psi _{-}}\right\rangle  \label{eq55}
\end{equation}

It is instructive to consider the particle velocity measured by physical
observers of the two sheets. For an observer of the $(+)$ sheet, any
particle confined to this sheet moves according to the velocity operator $%
V_{+}$ (notice the disappearance of the warp factor terms):

\begin{equation}
\left( V_{+}\right) _i=\left( {\frac{\partial H}{\partial \mathbf{p}_{+}}}%
\right) _i=\left( {{\
\begin{array}{cc}
{\gamma ^0\gamma ^i} & 0 \\
0 & 0
\end{array}
}}\right)  \label{eq56}
\end{equation}
Similarly any particle located in the $(-)$ sheet will be observed by local
observers as moving according to the velocity operator:

\begin{equation}
\left( V_{-}\right) _i=\left( {\frac{\partial H}{\partial \mathbf{p}_{-}}}%
\right) _i=\left( {{\
\begin{array}{cc}
0 & 0 \\
0 & {\gamma ^0\gamma ^i}
\end{array}
}}\right)  \label{eq57}
\end{equation}
Therefore, the global velocity operator of a particle unrestricted in its
motion can be written:

\begin{equation}
\left( V\right) _i=\frac 1{R_{+}}\left( V_{+}\right) _i+\frac 1{R_{-}}\left(
V_{-}\right) _i  \label{eq58}
\end{equation}
such that in the most general case, the global velocity reads:

\begin{equation}
\mathbf{v}=\mathbf{v}_{+}/R_{+}+\mathbf{v}_{-}/R_{-} \label{eq59}
\end{equation}
with
\begin{equation}
\mathbf{v}_{+}=\left\langle \psi \right| \mathbf{V}_{+}\left| \psi
\right\rangle =\left\langle {\psi _{+}}\right| \gamma ^0\gamma ^ie_i\left| {%
\psi _{+}}\right\rangle  \label{eq60}
\end{equation}
and similarly for $\mathbf{v}_{-}$.

We see that the particle velocity can be expressed as a combination
of the physical velocities of the particle, had it remained confined
in the $(+)$ or the $(-)$ sheet. As the particle oscillates from one
sheet to the other one with a time period $T$, the effective
distance $x$ that the particle is able to travel during one period
is given by (assuming a rectilinear motion):
\begin{equation}
x=\int_0^T{vdt}  \label{eq61}
\end{equation}
Let us define a starting point $x_{+}=0$ (or $x_{-}=0$) at $t=0$ in
the $(+)$ sheet (or resp. $(-)$ sheet). Therefore, an oscillating
particle detected at time $t=T$ in the $(+)$ sheet (or resp. $(-)$
sheet) is observed at a physical distance $x_{+}$ (or resp. $x_{-})$
from its starting point:

\begin{equation}
x_{+}=R_{+}x=\int\limits_0^T{\left( {v_{+}+\frac{R_{+}}{R_{-}}v_{-}}\right)
dt}\text{ and/or }x_{-}=R_{-}x=\int\limits_0^T{\left( {v_{-}+\frac{R_{-}}{%
R_{+}}v_{+}}\right) dt}  \label{eq62}
\end{equation}
As a consequence of the oscillatory motion, the particle is observed
at time $t=T$ and at a distance $x_{+}$ (or resp. $x_{-})$ as if it
had traveled at a fictitious mean velocity $\bar{v}_{+}$ (or
$\bar{v}_{-})$ from the perspective of an observer of the $(+)$
sheet (or resp. $(-)$ sheet):

\begin{equation}
\bar{v}_{+}=\frac 1Tx_{+}=R_{+}\frac 1T\int\limits_0^T{vdt}  \label{eq63}
\end{equation}
and/or
\begin{equation}
\bar{v}_{-}=\frac 1Tx_{-}=R_{-}\frac 1T\int\limits_0^T{vdt}  \label{eq64}
\end{equation}
with obviously: $\bar{v}_{+}\neq v_{+}$ and $\bar{v}_{-}\neq v_{-}$.

Therefore, in the most general case, the fictitious velocity
$(\bar{v}_{\pm })$ of an oscillating particle differs from that it
would have, had it remained confined in only one sheet (i.e. $v_{\pm
}$). To clarify this issue, let us consider the case of an
unpolarized particle initially mainly localized in the $(+)$ sheet
and described by the wave function
\begin{equation}
\psi =\frac 1{2\sqrt{V}}\left( {u_{1/2}e^{ip\cdot x}e^{-iE_pt}-\tilde{u}%
_{1/2}e^{ip\cdot x}e^{-i\tilde{E}_pt}}\right) +\frac 1{2\sqrt{V}}\left( {%
u_{-1/2}e^{ip\cdot x}e^{-iE_pt}-\tilde{u}_{-1/2}e^{ip\cdot x}e^{-i\tilde{E}%
_pt}}\right)  \label{eq65}
\end{equation}
Considering that the momentum $\mathbf{p}$ is along the $Oz$ axis, and
according to the convention used for the spinor determination, the particle
velocity becomes :
\begin{equation}
\mathbf{v}=ve_3=\left\langle \psi \right| V_3\left| \psi \right\rangle =\int
{\psi ^{\dagger }V_3\psi }d^3x  \label{eq66}
\end{equation}
We thus have
\begin{equation}
v=\frac 14\left\{ a-2b\cos \left( {E_p-\tilde{E}_p+\varphi }\right) t\right\}
\label{eq67}
\end{equation}
with:
\begin{equation}
a=u_{1/2}^{\dagger }V_3u_{1/2}+u_{-1/2}^{\dagger }V_3u_{-1/2}+\tilde{u}%
_{1/2}^{\dagger }V_3\tilde{u}_{1/2}+\tilde{u}_{-1/2}^{\dagger }V_3\tilde{u}%
_{-1/2}  \label{eq68}
\end{equation}
\begin{equation}
be^{i\varphi }=u_{1/2}^{\dagger }V_3\tilde{u}_{1/2}+u_{-1/2}^{\dagger }V_3%
\tilde{u}_{-1/2}\text{ and }be^{-i\varphi }=\tilde{u}_{1/2}^{\dagger
}V_3u_{1/2}+\tilde{u}_{-1/2}^{\dagger }V_3u_{-1/2}  \label{eq69}
\end{equation}
As a consequence, the mean fictitious velocity as seen by an observer of the
$(+)$ sheet becomes:
\begin{equation}
\bar{v}_{+}=R_{+}\frac 1T\int\limits_0^T{vdt}=\frac 14R_{+}a  \label{eq70}
\end{equation}
After tedious calculations, the fictitious particle velocity from the
perspective of an observer located in the $(+)$ sheet reads:
\begin{equation}
\bar{v}_{+}=R_{+}\frac pC\left( {E_p+m}\right) \left[ {\frac{\eta ^2}{R_{+}^2%
}+\frac{\left| {\vartheta _{1/2}}\right| ^2}{R_{-}^2}-\eta \xi ^2}\right]
+R_{+}\frac p{\widetilde{C}}\left( {\tilde{E}_p+m}\right) \left[ {\frac{%
\tilde{\eta}^2}{R_{+}^2}+\frac{\left| {\vartheta _{1/2}}\right| ^2}{R_{-}^2}-%
\tilde{\eta}\xi ^2}\right]  \label{eq71}
\end{equation}
For illustrative purposes, it is convenient to simplify this expression by
considering its non relativistic limit. Then assuming that $g<<m,$ $p$ and $%
p<<m$, one gets the more compact form
\begin{equation}
\overline{v}_{+}\cong \left\{ \frac \chi {2m}R_{+}+O[g]^2\right\}
p+O[p]^2=\frac 12\left( {1+\frac{R_{+}^2}{R_{-}^2}}\right) \frac{p_{+}}m
\label{eq72}
\end{equation}
with $\chi $ given by Eq.(23).

On the other hand, at the decoupling limit $g\to 0$ and assuming completely
localized particles in the $(+)$ sheet the velocity can be calculated to be
(from Eq.(44))

\begin{equation}
v_{+}=\frac 1{E_p}\frac p{R_{+}}=\frac{p_{+}}{E_p}  \label{eq73}
\end{equation}
At low impulsion, this expression conforms with the usual one as expected,
i.e. $v_{+}=p_{+}/m$. As a consequence, the expression (72) can be
conveniently rewritten as
\begin{equation}
\bar{v}_{+}={\frac 12\left( {1+\frac{R_{+}^2}{R_{-}^2}}\right) }v_{+}
\label{eq74}
\end{equation}
We see confirmed that an oscillating particle can travel between two
locations with a fictitious (but effective) speed $\bar{v}_{+}$ which
differs (in the general case where $R_{+}\neq R_{-}$) from that it would
have, had it remained in only one sheet (i.e. $v_{+}$ in the $(+)$ sheet).
If $R_{-}<<R_{+}$, the apparent particle velocity $\bar{v}_{+}$ can become
huge even if the particle velocity $v_{+}$ in the $(+)$ sheet remains
moderated.

A numerical example can be given for illustrative purposes. Let us assume a
warp factor ratio $R_{+}/R_{-}=250$ and a particle initially located in the $%
(+)$ sheet with a non relativistic velocity $v_{+}=20$ $km\cdot
s^{-1}$. Note that since we have $v_{+}/v_{-}=R_{-}/R_{+}$ in the
non relativistic limit (from Eq.(73)), we see immediately that if a
particle initially in the $(+)$ sheet reaches the $(-)$ sheet and
stays there, its physical
velocity as measured by an observed of the $(-)$ sheet will not be $20$ $%
km\cdot s^{-1}$ but $5000$ $km\cdot s^{-1}$. Although this is an important
velocity increase, this value is still a non relativistic one. At this
stage, it is important to stress that the equations of the physical
velocities $v_{+}$ (see Eq.(73)) and $v_{-}$ imply that $v_{\pm }<1$
whatever $R_{+},$ $R_{-},$ $p,$ $m$. Therefore, even if the particle moves
faster in the $(-)$ sheet, it can never exceed the light velocity. The laws
of special relativity are safe in the present approach. In addition, from
Eq.(45) we note that $x_{+}/x_{-}=R_{+}/R_{-}$. In the present example, this
means that the distances are $250$ times shorter in the $(-)$ sheet than in the $%
(+)$ sheet. Let us consider a particle initially localized in the $(+)$
sheet and transferred into the $(-)$ sheet. The particle covers a distance $%
x_{-}$ during a time $t$ and goes back in the $(+)$ sheet. For an observer
of the $(+)$ sheet, the detected particle has apparently moved a distance $%
x_{+}$ during a time $t$, such that its apparent (but fictitious) velocity $%
\overline{v}_{+}=x_{+}/t$ or $\overline{v}_{+}=v_{+}\left(
R_{+}/R_{-}\right) ^2$. The latter expression is the consequence of
the shortened distance and increased proper velocity in $(-)$ sheet.
The reason why this expression differs from Eq.(74) is the result of
the oscillating behavior of the particle. Indeed, on the average,
the oscillating particle spends as much time in sheet $(-)$ than in
sheet $(+)$. The fictitious velocity $\bar{v}_{+}$ in Eq.(74) is
then an average velocity between the real velocity in $(+)$ sheet
and the fictitious velocity, had it remained in the $(-)$ sheet
only.

For a proper velocity $v_{+}$ of $20$ $km\cdot s^{-1}$ in the $(+)$ sheet,
Eq.(74) shows that the fictitious velocity $\bar{v}_{+}$ (calculated by an
observer located in the $(+)$ sheet) is about $2$ times the light velocity.
Although the proper velocity of the particle remains non relativistic in
both sheets, its apparent velocity from the perspective of an observer of
the $(+)$ sheet exceeds now the light speed.

In Ref. [13], it was suggested that the homogeneity of the universe
by the time of nucleosynthesis might be explained by particles
motions in an asymmetric two-branes system (one of which being a
''hidden'' brane). According to the authors, an impulse originating
on one brane of the system can take a shortcut through the other
brane and affect our brane at a point outside the conventional
causal horizon. The present paper tries to go further on those
aspects by using a very simple although realistic quantum mechanical
model. Although our approach is radically different from that of
Ref. [13], our results share obvious similarities (e.g. the
particles confined in the other sheet are invisible to us, the
oscillating particles can reach distant point outside their
''naive'' horizon...). Hence, it is believed that there may
generically exist non-inflationary solution to the horizon problem
in theories with extra-dimensions (be they continuous or
discontinuous).

\section{Two-sheeted electromagnetic field and Pauli equation}

\subsection{Introduction of the Electromagnetic field}

To account for the discrete structure of the bulk, the usual $U(1)$ gauge
field must be substituted by an extended $U(1)\otimes U(1)$ gauge field such
that
\begin{equation}
G=\left[
\begin{array}{cc}
\mathbf{1}_{4\times 4}\exp (-iq\Lambda _{+}) & 0 \\
0 & \mathbf{1}_{4\times 4}\exp (-iq\Lambda _{-})
\end{array}
\right]  \label{eq75}
\end{equation}
We look for an appropriate gauge such that ${\not{D}}_A\rightarrow {\not{D}}+%
\not{A}$ with the following rule of transformation
\begin{equation}
\not{A}^{\prime }=G\not{A}G^{\dagger }+G\left[ {\not{D}},G^{\dagger }\right]
\label{eq76}
\end{equation}
A convenient choice is (see Refs. [9-12,16])
\begin{equation}
\not{A}=\left[
\begin{array}{cc}
iq\frac 1{R_{+}}\gamma ^\mu A_\mu ^{+} & \gamma ^5\chi \\
\gamma ^5\chi ^{\dagger } & iq\frac 1{R_{-}}\gamma ^\mu A_\mu ^{-}
\end{array}
\right]  \label{eq77}
\end{equation}
where $\gamma ^\mu $ and $\gamma ^5$ are the usual Dirac matrices. Choosing $%
\chi =0$ whatever the gauge choice can be done only if $\Lambda _{+}=\Lambda
_{-}=\Lambda $.

In addition, the above definitions impose the following gauge
transformations
\begin{equation}
A_\mu ^{\pm }=A_\mu ^{\pm }+\partial _\mu \Lambda  \label{eq78}
\end{equation}
On the two sheets live the distinct $A_{+}$ and $A_{-}$ fields. Each
spacetime sheet possesses its own current and charge density distribution as
sources of the local electromagnetic fields. The off-diagonal term has been
set equals to zero. This free choice allows a further simplification of the
model. If this term is different from zero, it leads to a coupling between
the two photons fields such that each charged particle becomes sensitive to
the electromagnetic fields of both sheets irrespective of its localization
in the bulk. With the present choice, the electromagnetic field of a sheet
couples only with the particles belonging to the same sheet.

\subsection{Derivation of the Pauli equation}

Let us derive the non relativistic limit of the two-sheeted Dirac equation
for the metric field (3). We first introduce the gauge contributions $\not%
{A} $ of both sheets into the two-sheeted Dirac equation following the
standard procedure
\begin{equation}
\left( i{\not{D}}_A-m\right) \Psi =0  \label{eq79}
\end{equation}
such that
\begin{equation}
{\not{D}}_A=\Gamma ^0\left( \partial _0+iq\widehat{A}_0\right) +\frac
1R\Gamma ^\eta \left( \partial _\eta +iq\widehat{A}_\eta \right) +g\Gamma
^5+g\Delta  \label{eq80}
\end{equation}
One can also write
\begin{equation}
i\partial _0\Psi =-i\Gamma ^0\frac 1R\Gamma ^\eta \left( \partial _\eta +iq%
\widehat{A}_\eta \right) \Psi -ig\Gamma ^0\Gamma ^5\Psi -ig\Gamma ^0\Delta
\Psi +m\Gamma ^0\Psi +q\widehat{A}_0\Psi  \label{eq81}
\end{equation}
where
\begin{equation}
\widehat{A}_\mu =\left[
\begin{array}{cc}
A_\mu ^{+} & 0 \\
0 & A_\mu ^{-}
\end{array}
\right]  \label{eq82}
\end{equation}
When $m$ is large compared with the kinetic energy, the most rapid time
dependence is in the factor $\exp (\pm imt)$. For a free positive energy
particle, and for small kinetic and electromagnetic energies, we may
therefore seek a solution of the form $\Psi =\psi e^{-imt}$ with
\begin{equation}
\psi =\left[
\begin{array}{c}
\chi _{+} \\
\theta _{+} \\
\chi _{-} \\
\theta _{-}
\end{array}
\right]  \label{eq83}
\end{equation}
where $\chi _{+}$, $\theta _{+}$, $\chi _{-}$, $\theta _{-}$ are
two-component spinors. One can then write

\begin{equation}
i\partial _0\chi _{+}=qA_0^{+}\chi _{+}-i(1/R_{+})\sigma _\eta \left(
\partial _\eta +iqA_\eta ^{+}\right) \theta _{+}-ig\left( \theta _{+}-\theta
_{-}\right) -ig\left( 3/2\right) \left\{ \left( R_{+}-R_{-}\right)
/R_{+}\right\} \theta _{+}  \label{eq84}
\end{equation}
\begin{equation}
i\partial _0\chi _{-}=qA_0^{-}\chi _{-}-i(1/R_{-})\sigma _\eta \left(
\partial _\eta +iqA_\eta ^{-}\right) \theta _{-}+ig\left( \theta _{+}-\theta
_{-}\right) -ig\left( 3/2\right) \left\{ \left( R_{-}-R_{+}\right)
/R_{-}\right\} \theta _{-}  \label{eq85}
\end{equation}
\begin{equation}
i\partial _0\theta _{+}=-i(1/R_{+})\sigma _\eta \left( \partial _\eta
+iqA_\eta ^{+}\right) \chi _{+}+qA_0^{+}\theta _{+}+ig\left( \chi _{+}-\chi
_{-}\right) -2m\theta _{+}+ig\left( 3/2\right) \left\{ \left(
R_{+}-R_{-}\right) /R_{+}\right\} \chi _{+}  \label{eq86}
\end{equation}
\begin{equation}
i\partial _0\theta _{-}=-i(1/R_{-})\sigma _\eta \left( \partial _\eta
+iqA_\eta ^{-}\right) \chi _{-}+qA_0^{-}\theta _{-}-ig\left( \chi _{+}-\chi
_{-}\right) -2m\theta _{-}+ig\left( 3/2\right) \left\{ \left(
R_{-}-R_{+}\right) /R_{-}\right\} \chi _{-}  \label{eq87}
\end{equation}
In addition, when $p$ is much smaller than $m$, $\theta _{+}$ and $\theta
_{-}$ become tiny in comparison with $\chi _{+}$ and $\chi _{-}$. For small
electromagnetic and kinetic energies, one then get the following expressions
\begin{equation}
\theta _{+}\approx -i(1/R_{+})\frac 1{2m}\sigma _\eta \left( \partial _\eta
+iqA_\eta ^{+}\right) \chi _{+}+i\frac g{2m}\left( \chi _{+}-\chi
_{-}\right) +i\frac g{2m}\left( 3/2\right) \left\{ \left( R_{+}-R_{-}\right)
/R_{+}\right\} \chi _{+}  \label{eq88}
\end{equation}
\begin{equation}
\theta _{-}\approx -i(1/R_{-})\frac 1{2m}\sigma _\eta \left( \partial _\eta
+iqA_\eta ^{-}\right) \chi _{-}-i\frac g{2m}\left( \chi _{+}-\chi
_{-}\right) +i\frac g{2m}\left( 3/2\right) \left\{ \left( R_{-}-R_{+}\right)
/R_{-}\right\} \chi _{-}  \label{eq89}
\end{equation}

In order to give these equations a more conventional form, vectors must be
used instead of covariant or contravariant terms. The procedure, which is
similar to that used in cosmology [17], can be summarized as follows. For a
metric field given by $g_{ij}=R^2\delta _{ij}$ $\left( {i,j=1,2,3}\right) $,
we can define an orthonormal basis:
\begin{equation}
\mathbf{e}_i=\frac{\mathbf{g}_i}R=R\mathbf{g}^i  \label{eq90}
\end{equation}
such that $g_{ij}=\mathbf{g}_i\cdot \mathbf{g}_j=g_{ji}$. What are usually
called the components of a vector $\mathbf{a}$ in elementary treatments are
neither the covariant components $a_i$ nor the contravariant components $a^i$%
, but the ``ordinary'' components:
\begin{equation}
\overline{a}_i=\mathbf{a}\cdot \mathbf{e}_i=Ra^i=R^{-1}a_i  \label{eq91}
\end{equation}
Moreover, to be consistent, vectors and operators relative to a specific
sheet have to be expressed using the metric of the corresponding sheet, i.e.
following Ref. [17]

\begin{equation}
\mathbf{\nabla }_{\pm }=\sum\limits_i\mathbf{e}_i{\frac 1{R_{\pm }}\frac
\partial {\partial x^i}}=\frac i{R_{\pm }}\sum\limits_i\mathbf{p}%
_i=i\sum\limits_i\mathbf{p}_{\pm ,i}\text{ where }\mathbf{p}_i=-i\mathbf{e}%
_i\frac \partial {\partial x^i}\text{ and }\mathbf{p}_{\pm ,i}=-i\mathbf{e}%
_i\frac \partial {\partial x_{\pm }^i}=\frac 1{R_{\pm }}\mathbf{p}_i
\label{eq92}
\end{equation}
and similarly, magnetic vector potentials and magnetic fields are given by
\begin{equation}
\bar{A}_{\pm ,i}=R_{\pm }A_{\pm }^i=R_{\pm }^{-1}A_{\pm ,i}\text{ and
similarly for }\bar{B}_{\pm ,j}  \label{eq93}
\end{equation}
Note that we have introduced the ``global'' vector $\mathbf{p}$ exactly as
we did for the Dirac equation (Eq.(45)).

By setting $\Phi _{\pm }$ and $\mathbf{A}_{\pm }$ for the usual electric and
magnetic potential, it can be easily shown that the Pauli equation reads

\begin{equation}
H\varphi =i\partial _0\varphi \text{ with }\varphi =\left[ {{\
\begin{array}{c}
{\chi _{+}} \\
{\chi _{-}}
\end{array}
}}\right]  \label{eq94}
\end{equation}
where we have used the fact that
\begin{equation}
\sigma _\eta \sigma _\nu \overline{\left( \mathbf{\nabla }_{\pm }-iq\mathbf{A%
}_{\pm }\right) }_\eta \overline{\left( \mathbf{\nabla }_{\pm }-iq\mathbf{A}%
_{\pm }\right) }_\nu =\left( \mathbf{\nabla }_{\pm }-iq\mathbf{A}_{\pm
}\right) ^2+q\mathbf{\sigma \cdot B}_{\pm }  \label{eq95}
\end{equation}
The resulting Hamiltonian can be written as the following sum
\begin{equation}
H=(H_k+H_m+H_p)\mathbf{+(}H_c+H_{cm}+H_{cp})  \label{eq96}
\end{equation}
where
\begin{eqnarray}
H_k=-\frac 1{2m}\left[
\begin{array}{cc}
\left( \mathbf{\nabla }_{+}-iq\mathbf{A}_{+}\right) ^2 & 0 \\
0 & \left( \mathbf{\nabla }_{-}-iq\mathbf{A}_{-}\right) ^2
\end{array}
\right]  \label{eq97}
\end{eqnarray}
\begin{eqnarray}
H_m=-\frac q{2m}\left[
\begin{array}{cc}
\mathbf{\sigma \cdot B}_{+} & 0 \\
0 & \mathbf{\sigma \cdot B}_{-}
\end{array}
\right]  \label{eq98}
\end{eqnarray}
\begin{eqnarray}
H_p=\left[
\begin{array}{cc}
q\Phi _{+} & 0 \\
0 & q\Phi _{-}
\end{array}
\right]  \label{eq99}
\end{eqnarray}
\begin{equation}
H_c=\frac{g^2}m\left[
\begin{array}{cc}
1+\left( 3/2\right) \left( \frac{R_{+}-R_{-}}{R_{+}}\right) +(9/8)\left(
\frac{R_{+}-R_{-}}{R_{+}}\right) ^2 & -1+\left( 3/4\right) \frac{\left(
R_{+}-R_{-}\right) ^2}{R_{+}R_{-}} \\
-1+\left( 3/4\right) \frac{\left( R_{+}-R_{-}\right) ^2}{R_{+}R_{-}} &
1+\left( 3/2\right) \left( \frac{R_{-}-R_{+}}{R_{-}}\right) +(9/8)\left(
\frac{R_{-}-R_{+}}{R_{-}}\right) ^2
\end{array}
\right]  \label{eq100}
\end{equation}
\begin{eqnarray}
H_{cm}=i\frac{gq}{2m}\left[
\begin{array}{cc}
0 & \mathbf{\sigma \cdot }\left\{ \mathbf{A}_{+}-\mathbf{A}_{-}\right\} \\
-\mathbf{\sigma \cdot }\left\{ \mathbf{A}_{+}-\mathbf{A}_{-}\right\} & 0
\end{array}
\right]  \label{eq101}
\end{eqnarray}
\begin{eqnarray}
H_{cp}=\frac g{2m}\left[
\begin{array}{cc}
0 & \mathbf{\sigma }\cdot \left\{ \mathbf{\nabla }_{+}-\mathbf{\nabla }%
_{-}\right\} \\
-\mathbf{\sigma }\cdot \left\{ \mathbf{\nabla }_{+}-\mathbf{\nabla }%
_{-}\right\} & 0
\end{array}
\right]  \label{eq102}
\end{eqnarray}

This two-sheeted Pauli equation is very similar to that derived in Refs [9]
and [10] except for the gradient operator which is distinct in the two
sheets as a consequence of the two metric fields. It is worth being noticed
that at the limit of decoupling, we get two times the usual Pauli equation
as expected. Therefore, at the limit of small g, the difference between the
two-sheeted Pauli equation and the usual one are not expected to be
significant.

The first three terms of the Hamiltonian correspond to the classical
contributions of the ''one-sheeted'' Pauli's equation. $H_k$ is the
kinetic term whereas $H_m$ and $H_p$ relate to the magnetic and
coulomb terms respectively. The last three terms correspond to new
predictions of the model. The term $H_c$ behaves as a constant
coupling between the two sheets. This term is responsible for the
spontaneous particle oscillations studied previously in the
relativistic limit. $H_{cp}$ and $H_{cm}$ introduce the geometrical
coupling between the sheets through kinetic and magnetic terms. It
is worth stressing that the coupling which involves the kinetic
mixing disappears if the two sheets have the same warp factors
whereas $H_{cm}$ does not.

\subsection{Hyperfast velocities in the non relativistic limit}

As previously, three different velocity operators can be defined : $\mathbf{V%
}_{+}$ ($\mathbf{V}_{-}$) which corresponds to the velocity of a confined
particle in the $(+)$ (resp.$(-)$) sheet and the global velocity operator $%
\mathbf{V}$. These operators read

\begin{equation}
\mathbf{V}_{+}=\frac{\partial H}{\partial \mathbf{p}_{+}}=\left[
\begin{array}{cc}
\frac 1m\left( \mathbf{p}_{+}-q\mathbf{A}_{+}\right) & \frac{ig}{2m}\mathbf{%
\sigma } \\
-\frac{ig}{2m}\mathbf{\sigma } & 0
\end{array}
\right]  \label{eq103}
\end{equation}
and
\begin{equation}
\mathbf{V}_{-}=\frac{\partial H}{\partial \mathbf{p}_{-}}=\left[
\begin{array}{cc}
0 & -\frac{ig}{2m}\mathbf{\sigma } \\
\frac{ig}{2m}\mathbf{\sigma } & \frac 1m\left( \mathbf{p}_{-}-q\mathbf{A}%
_{-}\right)
\end{array}
\right]  \label{eq104}
\end{equation}
and
\begin{eqnarray}
\mathbf{V} &=&\frac{\partial H}{\partial \mathbf{p}}=\left[
\begin{array}{cc}
\frac 1{R_{+}}\frac 1m\left( \frac 1{R_{+}}\mathbf{p}-q\mathbf{A}_{+}\right)
& \frac{ig}{2m}\mathbf{\sigma }\cdot \left\{ \frac 1{R_{+}}-\frac
1{R_{-}}\right\} \\
-\frac{ig}{2m}\mathbf{\sigma }\cdot \left\{ \frac 1{R_{+}}-\frac
1{R_{-}}\right\} & \frac 1{R_{-}}\frac 1m\left( \frac 1{R_{-}}\mathbf{p}-q%
\mathbf{A}_{-}\right)
\end{array}
\right]  \nonumber  \label{43} \\
&=&\frac 1{R_{+}}\frac{\partial H}{\partial \mathbf{p}_{+}}+\frac 1{R_{-}}%
\frac{\partial H}{\partial \mathbf{p}_{-}}=\frac 1{R_{+}}\mathbf{V}%
_{+}+\frac 1{R_{-}}\mathbf{V}_{-}  \label{eq105}
\end{eqnarray}
We point out the existence of a non diagonal components of the velocity
operator. However, since this term is proportional to $g$ we can ignore it
for the moment and in the forthcoming calculations to concentrate only on
the effect of the diagonal terms in $\mathbf{V}$.

\subsection{Illustrative case}

In the following, we neglect the terms $g^2$ of the hamiltonian in front of
the $g$ terms. In this way we underline the contribution of $H_{cm}$ and $%
H_{cp}$. To illustrate the survivance of hyperfast displacement in the non
relativistic limit, it is convenient to consider the case of a particle
initially located in the first sheet and embedded in a region of constant
curlless magnetic vector potential. For simplicity reasons, let us consider
a neutron-like particle, i.e. a chargeless particle with a magnetic moment.
We do not consider the complications arising from the anomalous magnetic
moment of this particle. Instead, we assume that it is identical to that of
an electron. In absence of any magnetic field nor scalar potential and by
neglecting the $g^2$ terms in front of the $g$ terms, the Hamiltonian reads
\begin{equation}
H=\left[
\begin{array}{cc}
K_{+} & -i\alpha \mathbf{\sigma \cdot P} \\
i\alpha \mathbf{\sigma \cdot P} & K_{-}
\end{array}
\right]  \label{eq106}
\end{equation}
with $\alpha =g/(2m)$ and
\begin{equation}
K_{\pm }=\frac 1{R_{\pm }^2}\frac{\mathbf{p}^2}{2m}  \label{eq107}
\end{equation}
and
\begin{equation}
\mathbf{P}=e\mathbf{A+}\left\{ \frac 1{R_{+}}-\frac 1{R_{-}}\right\} \mathbf{%
p}  \label{eq108}
\end{equation}
In the following, one considers that $\mathbf{A}$ and $\mathbf{p}$ are
colinear. The eigenvectors of the above hamiltonian are
\begin{equation}
u_{\pm ,\lambda }=\frac 1{\sqrt{N_{\pm }}}\left[
\begin{array}{c}
\left( E_{\pm }-K_{-}\right) \chi _\lambda \\
i\alpha \mathbf{\sigma \cdot P}\chi _\lambda
\end{array}
\right]  \label{eq109}
\end{equation}
with the corresponding eigenvalues
\begin{equation}
E_{\pm }=\frac 12\left( K_{+}+K_{-}\pm \sqrt{\left( K_{+}-K_{-}\right)
^2+4\alpha ^2P^2}\right)  \label{eq110}
\end{equation}
$\chi _\lambda $ is a spinor where $\lambda =\pm 1/2$ and stands for both
spin states and $N_{\pm }=(E_{\pm }-K_{-})\left( 2E_{\pm
}-K_{+}-K_{-}\right) $.

It is not difficult to convince oneself that the particle ability of
reaching the second spacetime sheet still survives in the non-relativistic
limit. The wave function assuming that the particle is initially ($t=0$)
located in the $(+)$ sheet with some polarization state $%
(n_{1/2}^2-n_{-1/2}^2)/(n_{1/2}^2+n_{-1/2}^2)$ is
\begin{equation}
\psi =\frac 1{\sqrt{V}}\left\{ n_{1/2}\left(
au_{+,1/2}e^{-iE_{+}t}+bu_{-,1/2}e^{-iE_{-}t}\right) +n_{-1/2}\left(
au_{+,-1/2}e^{-iE_{+}t}+bu_{-,-1/2}e^{-iE_{-}t}\right) \right\} e^{ip\cdot x}
\label{eq111}
\end{equation}
with $n_{1/2}^2+n_{-1/2}^2=1$ and
\begin{equation}
a=\sqrt{\frac{N_{+}}{N_{+}+N_{-}}}\text{and }b=\sqrt{\frac{N_{-}}{N_{+}+N_{-}%
}}  \label{eq112}
\end{equation}
The probability $P$ to find the particle in the second sheet can be
trivially calculated
\begin{equation}
P=\frac 1{1+\kappa ^2}\sin ^2\left\{ (1/2)(E_{+}-E_{-})t\right\}
\label{eq113}
\end{equation}
with
\begin{equation}
\kappa =\frac{K_{+}-K_{-}}{2\alpha P}  \label{eq114}
\end{equation}
Without going further, it can already be noticed that if $A=0$
\begin{equation}
\kappa =\frac{p_{+}}{2g}\left( 1+\frac{R_{+}}{R_{-}}\right)  \label{eq115}
\end{equation}
and
\begin{equation}
E_{+}-E_{-}=\frac{p_{+}^2}{2m}\left| 1-\frac{R_{+}}{R_{-}}\right| \sqrt{%
\left( 1+\frac{R_{+}}{R_{-}}\right) ^2+\frac{4g^2}{p_{+}^2}}  \label{eq116}
\end{equation}
with $p_{+}=p/R_{+}$. In absence of confining effect, the particle
oscillates between the two four-dimensional sections provided that $%
p_{+}\neq 0$. However, if $p_{+}$ becomes larger than $g$, $P$ quickly drops
to zero and the spontaneous oscillations are strongly suppressed.

Returning back to the general case where $A\neq 0$ and $p_{+}\neq 0$, it is
interesting to calculate the particle velocity. Assuming $\mathbf{p}$ and $%
\mathbf{A}$ are oriented along the $Oz$ axis, the global particle velocity $%
v $ is given by
\begin{eqnarray}
v &=&v_z=\int \psi ^{\dagger }V_z\psi dV  \label{eq117} \\
&=&\frac 1{R_{+}^2}\frac pm\cos ^2\left\{ (1/2)(E_{+}-E_{-})t\right\} +\frac
1{R_{-}^2}\frac pm\left( \frac{\frac{R_{-}^2}{R_{+}^2}\left(
K_{+}-K_{-}\right) ^2+4\alpha ^2P^2}{\left( K_{+}-K_{-}\right) ^2+4\alpha
^2P^2}\right) \sin ^2\left\{ (1/2)(E_{+}-E_{-})t\right\}  \nonumber
\end{eqnarray}
Since the particle oscillates between the two sheets, the velocity exhibits
also an oscillatory behavior. The fictitious velocity $\overline{v}_{+}$
from the point of view of an observer in the sheet $(+)$, after a period $T=%
\frac{2\pi }{(E_{+}-E_{-})}$ is:
\begin{equation}
\overline{v}=\frac{R_{+}}T\int_0^Tvdt  \label{eq118}
\end{equation}
and therefore:
\begin{equation}
\overline{v}_{+}=\frac 1{R_{+}}\frac pm\frac 12\left\{ 1+\frac{\left(
K_{+}-K_{-}\right) ^2+4\alpha ^2P^2\frac{R_{+}^2}{R_{-}^2}}{\left(
K_{+}-K_{-}\right) ^2+4\alpha ^2P^2}\right\}  \label{eq119}
\end{equation}
From Eqs.(107) and (108) and for a large enough $A$, the previous expression
simplifies further:
\begin{equation}
\overline{v}_{+}\approx \frac{v_{+}}2\left\{ 1+\frac{R_{+}^2}{R_{-}^2}%
\right\}  \label{eq120}
\end{equation}
with $v_{+}=p_{+}/m$ the usual particle velocity had it remained confined in
the $(+)$ sheet. We recover the relation already derived from the
relativistic approach at the limit of low velocities.

Again, we see that if $R_{+}>>R_{-}$, the particle moves with an
apparent velocity that can exceed the velocity $v_{+}$ this particle
would have, had it remained in the $(+)$ sheet. The novelty in
comparison with the result (74) is that the particle oscillates
through the application of a magnetic potential and therefore this
result suggest that the model could perhaps be experimentally
investigated (this possibility and the properties of $H_{cm}$ have
been explored more in Refs. [11,12]).

We stress that the results of this paper have been obtained for a
free particle, e.g. by assuming that the magnetic field, the scalar
potential and any other environmental contribution can be neglected
(excepted for a constant curlless magnetic vector potential as
discussed previously). This condition is obviously a very
restrictive one and in most cases, the suppression of these terms
could be hardly achieved. In Ref. [12], it was demonstrated that any
diagonal term in the Hamiltonian strongly suppress the particle
oscillations. In fact, it is not difficult to demonstrate that the
more energetic the particle is, the more the oscillations are
suppressed. Even the gravitational potential, whose contribution
modifies the particle energy, will affect the oscillations and
restricts the particle motion between the sheets. As demonstrated in
Ref. [12], our model suggests that any massive particle could be
spontaneously confined within the sheets (through environmental
interactions) without requiring any complementary scalar field or
repulsive gravity [12]. Since no particle disappearance has been
noted to date, it is very likely that the degree of confinement is
very strong, and/or that the coupling constant $g$ is very small.
Therefore, it is expected that the phenomena described in this paper
will be hardly observed, especially if one takes into account the
fact that no current experimental setup is suitably designed for
searching for these phenomena. Any experiment aiming at demonstrated
the behavior predicted in the present paper will require very
particular conditions although they might not be completely out of
reach of our present technology [11,12].

\section{Conclusions}

In this paper, we have studied the quantum dynamics of spin half
particles in an asymmetrically two-sheeted spacetime. It was shown
that any free particle oscillates between the two sheets as a
consequence of the geometrical coupling along the discrete
extra-dimension. By oscillating, any massive particle is able to
travel between distant points which are normally outside its
four-dimensional horizon. The reason arises from the differential
warping which leads to reduced length scales and increased
velocities in one of the sheet.

\appendix
\section{}
The discrete derivative used in present and previous papers [10-12]
is defined as follows.

A compact oriented discretized dimension $Z_n=\left\{ s_i|i\in
\left\{ 0,...,n-1\right\} \right\} $ is considered, with $s_i$ are
each of the $n$ sites of $Z_n$ (see figure 1). $Z_n$ is invariant
under the cyclic group $Z/nZ$. $Z_n$ is oriented positively with
increasing values of site
index $i$. For each site $s_i$ of $Z_n$ it is possible to define a coordinate $%
\lambda (s_i)=\lambda _i=\delta i+q$. Note that both $\delta $ and
$q$ have the dimension of a length. $q$ is an arbitrary constant
(the presence of this term will be clarified shortly afterwards)
whereas $\delta $ is the distance between nearest-neighbor sites.

\begin{figure}[hp]
\centerline{\psfig{file=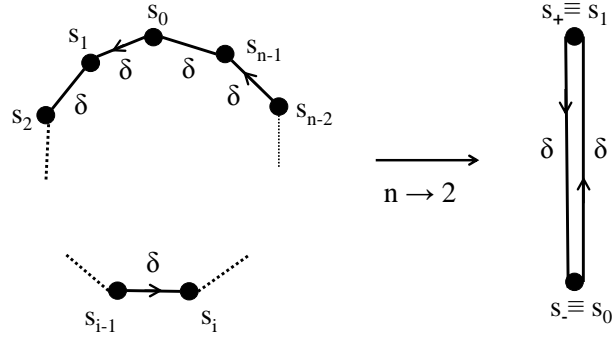,width=8cm,height=4.4cm}}
\vspace*{8pt} \caption{General polygonal representation of the
compact discretized $Z_n$ dimension (left) and of its $Z_2$ limit
(right). } \label{fig1}
\end{figure}

One then defines a positive restricted algebraic distance $d_a(s_i,s_j)$ on $%
Z_n$. That means that one goes from site $s_i$ to site $s_j$ by
imposing a positive direction, i.e.
\begin{equation}
d_a(s_i,s_j)=\left\{
\begin{array}{c}
\lambda _j-\lambda _i \\
\lambda _j-\lambda _i+\delta n
\end{array}
\right.
\begin{array}{c}
\text{if }j\geq i \\
\text{if }j<i
\end{array}
\label{A1}
\end{equation}
This imposes
\begin{equation}
d_a(s_{i-1},s_i)=\delta \text{ for }i\in \left\{ 1,...,n-1\right\}
\label{A2}
\end{equation}
and
\begin{equation}
d_a(s_{n-1},s_0)=\delta   \label{A3}
\end{equation}
with respect to the cyclic properties.

If one considers an arbitrary function $\phi (\lambda (s_i))=\phi
(\lambda _i)=\phi _i$, the  derivative $\partial _\lambda \phi
(\lambda _i)$ at each site $s_i$ is naturally given by:
\begin{equation}
\partial _\lambda \phi (\lambda _i)\equiv \frac{\phi (\lambda _i)-\phi
(\lambda _{i-1})}{d_a(s_{i-1},s_i)}\text{ for }i\in \left\{
1,...,n-1\right\}   \label{A4}
\end{equation}
with of course
\begin{equation}
\partial _\lambda \phi (\lambda _0)\equiv \frac{\phi (\lambda _0)-\phi
(\lambda _{n-1})}{d_a(s_{n-1},s_0)}  \label{A5}
\end{equation}

These expressions can be written in the usual form
\begin{equation}
\partial _\lambda \phi _i\equiv (1/\delta )\left( \phi _i-\phi _{i-1}\right)
\ \text{for\ }i\in \left\{ 1,...,n-1\right\}   \label{A6}
\end{equation}
and
\begin{equation}
\partial _\lambda \phi _0\equiv (1/\delta )\left( \phi _0-\phi _{n-1}\right)
\label{A7}
\end{equation}
By construction, the derivative is invariant through cyclic permutation and corresponds to the so-called ''$Z_n$%
-derivative''.

The ''$Z_2$-derivative'' is simply obtained by setting $n=2$ in
previous expressions. The result reads:
\begin{equation}
\partial _\lambda \phi _1=(1/\delta )\left( \phi _1-\phi _0\right) \text{
and }\partial _\lambda \phi _0=(1/\delta )\left( \phi _0-\phi
_1\right) \label{A8}
\end{equation}
Considering the following substitution $\left( \phi _0,\phi
_1\right) \rightarrow \left( \phi _{-},\phi _{+}\right) $ the
derivative can be expressed into the form
\begin{equation}
\partial _\lambda \phi _{\pm }=\pm (1/\delta )\left( \phi _{+}-\phi
_{-}\right)   \label{A9}
\end{equation}
In addition, by setting $q=-\delta /2,$ wet get $\lambda _{\pm }=\pm
\delta /2$ for the coordinate of each site.


\begin{thebibliography}{99}
\bibitem{1}  Th. Kaluza, ``Zum Unit\"{a}tsproblem der Physik'', Sitz.
Preuss. Akad. Wiss. Phys. Math. Kl. (1921) 966

\bibitem{2}  L. Randall, R. Sundrum, ``Large mass hierarchy from a small
extra dimension'', Phys. Rev. Lett. 83, 3370 (1999) (hep-ph/9905221)

\bibitem{3}  L. Randall, R. Sundrum, ``An alternative to compactification'',
Phys. Rev. Lett. 83, 4690 (1999) (hep-th/9906064)

\bibitem{4}  N. Arkani-Hamed, S. Dimopoulos, G. Dvali, N. Kaloper, ``Manyfold Universe'', JHEP 0012 (2000) 010 (hep-ph/9911386)

\bibitem{5}  C. Deffayet, J. Mourad, ``Multigravity from a discrete
extradimension'', Phys. Lett. B589 (2004) 48-58 (hep-th/0311124). C.
Deffayet, J. Mourad, ``Solutions of multigravity theories and
discretized brane worlds'', Class. Quant. Grav. 21 (2004) 1833-1848
(hep-th/0311125)

\bibitem{6} N. Arkani-Hamed, M.D. Schwartz, ``Discrete Gravitational
dimensions'', Phys. Rev. D 69, 104001 (2004) (hep-th/0302110)

\bibitem{7}  A. Connes, J. Lott, ``Particle models and non-commutative
geometry'', Nucl. Phys. B18 (Proc.Suppl.) (1990) 29-47

\bibitem{8}  N.A.Viet, K.C.Wali, ``Non-commutative geometry and a
Discretized Version of Kaluza-Klein theory with a finite field content'',
Int. J. Mod. Phys. A11 (1996) 533 (hep-th/9412220)

\bibitem{9}  F. Petit, M. Sarrazin, ``Quantum dynamics of massive particles
in a non-commutative two-sheeted space-time'', Phys. Lett. B612 (2005)
105-114 (hep-th/0409084)

\bibitem{10}  M. Sarrazin, F. Petit, ``Quantum dynamics of particles in a
discrete two-branes world model: Can matter particles exchange occur between
branes?'', Acta Phys. Polon. B36 (2005) 1933-1950 (hep-th/0409083)

\bibitem{11}  M. Sarrazin, F. Petit, ``Artificially induced positronium
oscillations in a two-sheeted spacetime: consequences on the observed decay
processes'', Int. J. Mod. Phys. A21 (2006) 6303-6314 (hep-th/0505014)

\bibitem{12}  M. Sarrazin, F. Petit, ``Matter localization and resonant
deconfinement in a two-sheeted spacetime'', Int. J. Mod. Phys. A22
(2007) 2629-2641 (hep-th/0603194)

\bibitem{13}  D.J.H. Chung, K. Freese, ``Can geodesics in extra dimensions
solve the cosmological horizon problem?'', Phys. Rev. D 62, 063513 (2000)
(hep-ph/9910235)

\bibitem{14}  S.L. Dubovsky, V.A. Rubakov, P.G. Tinyakov, ``Brane world:
disappearing massive matter'', Phys. Rev. D 62, 105011 (2000)
(hep-th/0006046)

\bibitem{15}  R. Gregory, V.A. Rubakov, S.M. Sibiryakov, ``Brane worlds: the
gravity of escaping matter'', Class. Quant. Grav. 17, 4437 (2000)
(hep-th/0003109)

\bibitem{16}  Lizzi, G. Mangano, G. Miele, ``Another Alternative to
Compactification: Noncommutative Geometry and Randall-Sundrum Models'', Mod.
Phys. Lett. A16 (2001) 1-8 (hep-th/0009180)

\bibitem{17}  S. Weinberg, ``Gravitation and Cosmology: Principles and
Applications of the General Theory of Relativity'', John Wiley {\&} Sons,
New York (1972)
\end{thebibliography}
\end{document}